\documentclass[journal, a4paper]{IEEEtran}

\usepackage[noadjust]{cite}      

\usepackage{graphicx}   

\usepackage{url}        
\usepackage{amsmath}
\usepackage{amsfonts}

\usepackage{pgfplots}
\usepackage{tikz}
\pgfplotsset{compat=newest}%1.12}
\usepackage{siunitx}
\usepgfplotslibrary{units} % Allows to enter the units nicely
\usepackage{physics}
\usepackage{blochsphere}
\usepackage{xfrac}
\usepackage{listings}
\usepackage{xcolor}
\usepackage[caption=false]{subfig}
\usepackage{booktabs}
\usepackage{array}

\usepackage{algorithm}
\usepackage[noend]{algpseudocode}
\pdfoutput=1

\IEEEoverridecommandlockouts
\begin{document}

\pgfmathdeclarefunction{gauss}{2}{%
  \pgfmathparse{sqrt(1/(#2*sqrt(2*pi)))*exp(-((x-#1)^2)/(4*#2^2))}%
}
\newcommand{\xcmd}[2]{\left(#1\left\vert\vphantom{#1#2}\right. \!#2\right)}
\renewcommand{\algorithmicrequire}{\textbf{Input:}}
\renewcommand{\algorithmicensure}{\textbf{Output:}}

% Define document title and author
	\title{Comparing weak and projective measurements for quantum state tomography of a single-qubit system}
	\author{Shoumik Chowdhury$^\ast$$^\dagger$
 	\thanks{\noindent \hspace*{-1.3em} $^\ast$ {\em Aditya Birla World Academy, Mumbai, MH 400007, India} \newline
     $^\dagger$ {\em Quantum Measurement and Control Laboratory, Tata Institute of Fundamental Research (TIFR), Mumbai, MH 400005, India} }
     \vspace{-0.5cm}
     }
	
	\maketitle
	\makeatletter
	\def\BState{\State\hskip-\ALG@thistlm}
	\makeatother
% Write abstract here
\begin{abstract}
	We explore the use of weak quantum measurements for single-qubit quantum state tomography processes. Weak measurements are those where the coupling between the qubit and the measurement apparatus is weak; this results in the quantum state being disturbed less than in the case of a projective measurement. We employ a weak measurement tomography protocol developed by Das and Arvind, which they claim offers a new method of extracting information from quantum systems. We test the Das-Arvind scheme for various measurement strengths, and ensemble sizes, and reproduce their results using a sequential stochastic simulation. Lastly, we place these results in the context of current understanding of weak and projective measurements.   
\end{abstract}

% Each section begins with a \section{title} command
\section{Introduction}
	% \PARstart{}{} creates a tall first letter for this first paragraph
	\PARstart{I}{n} quantum mechanics, the state of a system can be fully represented by its wave function. Knowledge of the wave function of a quantum system serves as a direct means to predict the outcome of a measurement on an observable of that system. The ability to accurately determine the wave function of a possibly unknown quantum state is, therefore, of fundamental importance. 
    
    However, as we know, it is no easy task to `measure' the state of a quantum system. The notion of measurement itself carries a different connotation in quantum mechanics than in classical physics. Measurements performed on quantum systems are invasive, irrevocably disturbing the system. We say that measurements cause backaction on the state being measured. For `projective' measurements in particular, the wave function appears to `collapse' into only one of the eigenstates of the initial superposition. It is observed that subsequent measurements on a `collapsed' state give the same measurement result. That is to say, no further information can be gained. This is none other than the famous `measurement problem' of quantum mechanics \cite{QCQI}. As the outcome of one single projective measurement cannot be accurately predicted, measurements must be performed on many identical copies of the quantum system, known as an {\em ensemble}. This process is known as quantum state tomography, where a series of measurements is performed on an ensemble in order to estimate the quantum state \cite{QST}. 
    
	As opposed to projective measurements, another class of generalized measurement is known as the `weak' measurement. Here, the coupling between the system and the measuring device is made weak, thus causing proportionately less backaction. The superposition state does not fully collapse into one of the eigenstates, but instead the system may be reused for subsequent measurements \cite{Barchielli,Barchielli2,Milburn}. Consequently, however, this means that only small amounts of information can be extracted from each measurement \cite{Todd}. It is important to understand however, that both weak and projective measurements fall under the more general class of Positive Operator Valued Measurements (POVMs), which are often interpreted as projective measurements on a higher-dimension Hilbert space  \cite{QCQI,Todd}.
    
    Several schemes have been proposed that make use of weak measurements for various applications \cite{Prugovecki,Prugovecki2,Diosi,Ali,Kitagawa}. Specifically, several examples of quantum state tomography using weak measurements can be found in Refs. \cite{Hofmann,Wu,Tanaka}. Most notably, however, weak continuous measurements are finding applications in quantum error correction \cite{Ahn} and for feedback in superconducting quantum circuits \cite{Vijay1,Vijay2} where projective measurements cause too much backaction on the system. 
   
Building upon these, this study aims to explore the apparent dichotomy between weak and projective measurements. We consider a scheme proposed by Das and Arvind, which uses weak measurements for quantum state tomography. Das and Arvind believe that their protocol can outperform `standard' quantum state tomography based on projective measurements \cite{Arvind}. This result is surprising, and thus we seek to reproduce their protocol and place it into the context of current knowledge about weak measurements. We test the scheme on known quantum states using quantum Bayesian statistics and use the expectation values to reconstruct the state. This process is then compared for both weak and projective measurements. 

This paper is organized as follows. In Section II, we review certain key concepts from quantum measurement theory. In Section III, we present the Das-Arvind scheme and our simulation used to test it for both weak and projective measurements. We present the results of these simulated measurements in Section IV. Section V compares the results to other related work on these protocols, particularly that in Gross et. al. \cite{Gross}. Lastly Section VI discusses future work.

% Background
\section{Background}

\subsection{The von Neumann Model of Measurement}
	
    Although the von Neumann model initially set out to describe projective measurements, it can easily be applied to weak measurements as well \cite{Duck}. Let us consider a generalized quantum system with initial state vector $\ket{\psi}$, which can be expressed as  
    \begin{equation}
     \ket{\psi} = \sum_{i = 1}^n c_i \ket{a_i}
    \end{equation}
 with complex coefficients $c_i$ and eigenstates $\ket{a_1}, \ket{a_2}, \ldots \ket{a_n}$. These are the {\em basis states} described by the von Neumann model. As we are working with qubits, we will consider the specific case of a quantum two-level system, characterized by a two-dimensional Hilbert space $\mathcal{H}_2$. We can label the basis states of this system  $\ket{0}$ and $\ket{1}$, which refer to the eigenstates of the z-component of spin, $\sigma_z$. Any general state in this system can then be written as
 \begin{equation}
 \ket{\psi} = \alpha \ket{0} + \beta \ket{1}
 \label{eq:qubit}
 \end{equation}
 
   By convention, the eigenvalues of this system are $+1$ and $-1$, respectively. We know from Born's rule that the probability of the measurement outcome $\ket{0}$ is $\abs{\alpha}^2$ and for outcome $\ket{1}$ is $\abs{\beta}^2$. In order to satisfy the normalization condition $\bra{\psi}\ket{\psi} = 1$, clearly $\abs{\alpha}^2 + \abs{\beta}^2 = 1$. 
   
   Let us now consider a measurement device with a continuous meter describing the variable $q$. We follow the description given in \cite{Arvind}, wherein the initial quantum state $\ket{\phi}$ for this device follows a Gaussian distribution, with a spread (i.e. standard deviation) $\Delta q$. The state, then, is given by
   \begin{equation}
 \ket{\phi} = \sqrt[4]{\frac{\epsilon}{2 \pi}} \int_{-\infty}^\infty dq\,e^{-\epsilon q^2 / 4} \ket{q}  
 \label{eq:apparatus}
 \end{equation}
 where $\epsilon = 1/(\Delta q)^2$. The `weakness' of the measurement is characterized by $1/\sqrt{\epsilon}$. We see that $\abs{\ket{\phi}}^2$ give a standard Gaussian probability density distribution corresponding to the likeliness of obtaining each meter reading $q$. Real measurement devices have characteristics that are modeled well by a Gaussian quantum state, and thus this choice of distribution is apt \cite{Mensky}. 
 
 After the measurement apparatus interacts with the system, the two wave functions $\ket{\psi}$ and $\ket{\phi}$ become entangled, and the resulting output state $\ket{\psi_{out}}$ can be given by
\begin{multline}
\ket{\psi_{out}} = \sqrt[4]{\frac{\epsilon}{2 \pi}} \int_{-\infty}^\infty dq\,\alpha\,e^{-\frac{\epsilon (q - 1)^2}{4}} \ket{0} \otimes \ket{q} \, + \\ 
\sqrt[4]{\frac{\epsilon}{2 \pi}} \int_{-\infty}^\infty dq\,\beta\,e^{-\frac{\epsilon (q + 1)^2}{4}} \ket{1} \otimes \ket{q}
\label{eq:psiout}
\end{multline}

\begin{figure}[h]
\centering
\resizebox {\columnwidth} {!} {
\begin{tikzpicture}
\begin{axis}[
  no markers, domain=-2.5:2.5, samples=100,
  axis lines=middle, xlabel=$q$, ylabel=$P(q)$,
  every axis y label/.style={at=(current axis.above origin),anchor=south},
  every axis x label/.style={at=(current axis.right of origin),anchor=west},
  height=5cm, width=12cm,
  xtick={-1,1}, ytick=\empty,
  enlargelimits=false, clip=false, axis on top,
  grid = major
  ]
  \node at (axis cs:0,-0.13) {0};
  \addplot [very thick,red!50!black] {gauss(-1,.20)};
  \addplot [very thick,red!50!black] {gauss(1,.20)};
  \node at (axis cs:1.5,1.4) {$P\!\xcmd{q}{\ket{0}}$};
  \node at (axis cs:-1.5,1.4) {$P\!\xcmd{q}{\ket{0}}$};
\end{axis}
\end{tikzpicture}}
\caption{Illustration of quantum measurement outcomes described in Eq. \eqref{eq:psiout}}
\label{fig:outcomes}
\end{figure}
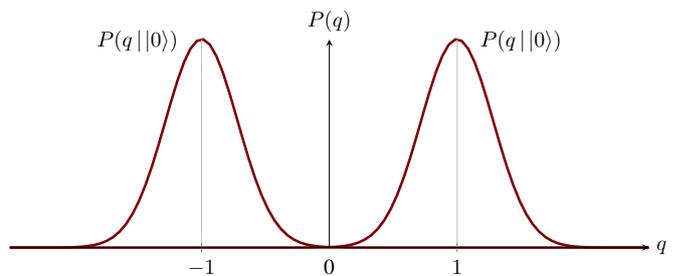

The state $\ket{\psi_{out}}$ corresponds to two Gaussian distributions centered around the respective eigenvalues of $\ket{0}$ and $\ket{1}$.  We must state explicitly, however, that the distributions given by  Eq. \eqref{eq:psiout} only represent the probability {\em amplitudes}. We can find the corresponding probability density functions by taking $\abs{\ket{\psi_{out}}}^2$, where the expectation value is $+1$ for the outcome $\ket{0}$ and $-1$ for the outcome $\ket{1}$. The conventional way to depict this is to plot the two probability density distributions as sketched in Fig. \ref{fig:outcomes}.

We use the notation P$\xcmd{\cdot}{\cdot}$ to express conditional probability. Thus, $P\!\xcmd{q}{\ket{0}}$ on Fig. \ref{fig:outcomes} should be read as `the probability of obtaining a meter reading $q$ given that the system is in the eigenstate $\ket{0}$.' From Bayes' theorem, and in particular its quantum counterpart, we can {\em update} the state based on the measurement result $q$ \cite{bayesianproof}. We know that
\begin{equation}
P\!\xcmd{\ket{0}}{q} = \frac{P(\ket{0}) \cdot P\!\xcmd{q}{\ket{0}}}{P(q)}
\label{eq:bayesian}
\end{equation}
where $P(\ket{0})$ is the {\em a priori} probability of the system collapsing into the $\ket{0}$ eigenstate and $P\!\xcmd{\ket{0}}{q}$ is the {\em a posteriori} probability after obtaining a pointer value $q$. The probability $P(q)$ is the combined probability distribution of obtaining a measurement value in either state, given simply by $P(q) = P(\ket{0}) \cdot P\!\xcmd{q}{\ket{0}} + P(\ket{1}) \cdot P\!\xcmd{q}{\ket{1}}$ 

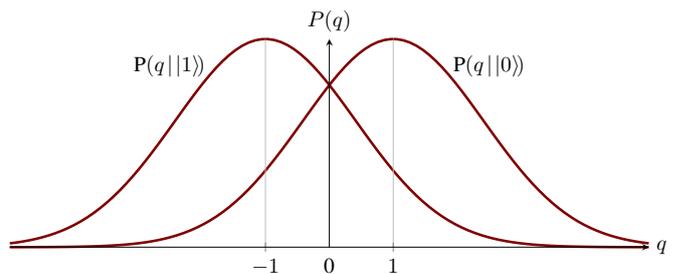
\begin{figure}[h]
\centering
\resizebox {\columnwidth} {!} {
\begin{tikzpicture}
\begin{axis}[
  no markers, domain=-5:5, samples=100,
  axis lines=middle, xlabel=$q$, ylabel=$P(q)$,
  every axis y label/.style={at=(current axis.above origin),anchor=south},
  every axis x label/.style={at=(current axis.right of origin),anchor=west},
  height=5cm, width=12cm,
  xtick={-1,1}, ytick=\empty,
  enlargelimits=false, clip=false, axis on top,
  grid = major
  ]
  \node at (axis cs:0,-0.055) {0};
  \addplot [very thick,red!50!black] {gauss(-1,1)};
  \addplot [very thick,red!50!black] {gauss(1,1)};
  \node at (axis cs:2.5,0.55) {P$(q\!\mid \!\ket{0}\!)$};
  \node at (axis cs:-2.5,0.55) {P$(q\!\mid \!\ket{1}\!)$};
\end{axis}
\end{tikzpicture}}
\caption{Illustration of a weak measurement, with a large initial spread $\Delta q$.}
\label{fig:outcomes_weak}
\end{figure}

Qualitatively, we see that the two distributions in Fig. \ref{fig:outcomes} do not overlap appreciably, and there is little ambiguity in the measurement result: a positive pointer value implies the outcome $\ket{0}$, whereas a negative value implies $\ket{1}$. If, however, we were to increase the initial spread of the pointer ($\Delta q$), the resultant Gaussians would look more like those in Fig. \ref{fig:outcomes_weak}. 

Here, there is an ambiguity in the measurement outcome, as the two Gaussians overlap significantly. Interestingly, a pointer value of 0 gives us no information about the quantum state but consequently causes no backaction, leaving the system unchanged! We will test the Das-Arvind protocol for various values of $\Delta q$ (i.e by varying $\epsilon$) to test whether weak measurements possess any advantages over projective measurements (which can be represented in our model as `strong' measurements where $\Delta q$ is very small).

\subsection{The Bloch Sphere and the Density Operator}

\begin{figure}[h]
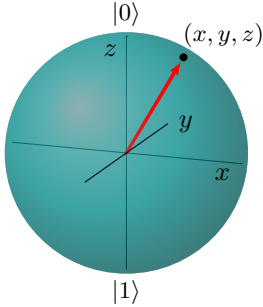

\centering
\begin{blochsphere}[radius=1.6cm,opacity=0.3,rotation=70, color=cyan]
\drawBall
%\drawBallGrid[style={opacity=0.2}]{30}{30}
%\drawAxis{90}{0}
%\drawAxis{90}{90}
\drawAxis{0}{0}
\drawAxis{90}{0}
\drawAxis{90}{90}

\labelLatLon{z-axis}{90}{0};
\labelLatLon{-z-axis}{-90}{0};
\labelLatLon{y-axis}{0}{0};
\labelLatLon{x-axis}{0}{90};
\node[left=6pt,below=1pt] at (z-axis) {$z$};
\node[above=2pt] at (z-axis) {{\small $\ket{0}$}};
\node[below=2pt] at (-z-axis) {{\small $\ket{1}$}};
\node[right=2pt] at (y-axis) {$y$};
\node[below=4pt,left=2pt] at (x-axis) {$x$};

\drawStatePolar[statewidth=0.7pt,statecolor=red,labelmark=true]{state}{30}{90}
\node[right=15pt,above=2pt] at (state) {\small $(x,y,z)$};
\end{blochsphere}
\caption{Bloch sphere representation of a qubit, with $\vec{a} = (x,y,z)$}
\label{fig:bloch}
\end{figure}

Recall that a Bloch sphere is a unit sphere where every point inside and on the surface of the sphere represents a unique quantum state \cite{QCQI}. Points on the surface represent pure states, whereas points within the sphere are indicative of mixed states; the latter exist as a mixture of states, and are often represented as density operators. The Bloch sphere is a useful tool to visualize qubits, and can be obtained by parametrizing Eq. \eqref{eq:qubit}, while still satisfying the normalization condition, as 
\begin{equation}
\ket{\psi} = \cos\frac{\text{\raisebox{-3pt}{$\theta$}}}{2}\ket{0} + e^{i\varphi} \sin\frac{\text{\raisebox{-3pt}{$\theta$}}}{2}\ket{1}
\label{eq:bloch}
\end{equation}
where the parameters $\theta$ and $\varphi$ specify the spherical coordinates on the surface of a unit sphere in $\mathbb{R}^3$. We see that the poles of this sphere correspond to the qubit eigenstates $\ket{0}$ and $\ket{1}$. `Pure states' are those whose coordinates lie on the surface of the Bloch sphere, and can be denoted by a position vector $\vec{a} = (x,y,z)$ = $(\sin\theta\cos\varphi, \sin\theta\sin\varphi, \cos\theta)$. An example of a state vector drawn on the Bloch sphere is given in Fig. \ref{fig:bloch}.

We can also draw a link between the Bloch sphere representation of a quantum state and the density operator, also known as a `density matrix'. Recall now that the density operator $\rho$ of a quantum state $\ket{\psi}$ is described most generally as 
\begin{equation}
\rho = \sum_{i} k_i \ket{\psi_i}\bra{\psi_i}
\label{eq:densitymatrix1}
\end{equation}

This is a compact representation for a quantum state and was devised by von Neumann in 1927 \cite{vonneumann}. As a shorthand to Eq. \eqref{eq:densitymatrix1}, $\rho$ can also be written as a matrix with elements $k_i$. The dimensions of this matrix correspond to the number of Hilbert dimensions of the quantum state: a wave function in $\mathcal{H}_n$ requires an $n\times n$ matrix. For the specific case of a qubit, we can use the Bloch sphere vector $\vec{a}$ to arrive at the density matrix, given that
\begin{equation}
\rho = \frac{1}{2} \big(I + x\sigma_x + y\sigma_y + z\sigma_z\big) = \frac{1}{2} \!
\begin{pmatrix}
1 + z & x - iy \\
x + iy & 1 - z
\end{pmatrix}
\label{eq:densitymatrix2}
\end{equation}
Here, $I$ is the identity matrix, and $\sigma_x, \sigma_y$, and $\sigma_z$ are the $2\times2$ Pauli matrices. 
\begin{equation*}
\sigma_x = 
\begin{pmatrix}
0 & 1 \\
1 & 0
\end{pmatrix} \qquad
\sigma_y = 
\begin{pmatrix}
0 & -i \\
i & 0
\end{pmatrix} \qquad
\sigma_z = 
\begin{pmatrix}
1 & 0 \\
0 & -1
\end{pmatrix}
\end{equation*} 

Using this new notation, let us write Eq. \eqref{eq:qubit} as a density matrix. The qubit state is given by $\ket{\psi} = \alpha \ket{0} + \beta \ket{1}$, and thus correspondingly $\bra{\psi} = \alpha^* \bra{0} + \beta^* \bra{1}$. Here, and throughout this study, the notation $^*$ signifies the complex conjugate. We see that \begin{align*}
\rho &= \abs{\alpha}^2\ket{0}\bra{0} + \beta^*\alpha\ket{0}\bra{1} + \alpha^*\beta\ket{1}\bra{0} + \abs{\beta}^2\ket{1}\bra{1} \\ 
&= \begin{pmatrix}
\abs{\alpha}^2 & \beta^*\alpha \\
\alpha^*\beta & \abs{\beta}^2
\end{pmatrix} 
\end{align*}
Clearly, the diagonal elements of $\rho$ give the probabilities of obtaining the outcomes $\ket{0}$ and $\ket{1}$, respectively. Thus, the normalization condition for a $2\times2$ density matrix is that its trace sums to 1. 

\subsection{Simulating weak measurements using quantum trajectories}

The action of weak measurements on a quantum state can be demonstrated using a quantum trajectory. Studying this example will help introduce several key methods that are useful for simulating quantum state tomography. If a series of weak measurements are performed on a quantum state, each measurement will cause some amount of backaction on the system. We can describe the discrete time evolution for a state $\rho$ as a function $\rho(t)$, corresponding to successive measurements being made on the state. This evolution is known as a quantum trajectory and was first introduced as a theoretical tool to study open quantum systems \cite{Carmichael,Gardiner,Dalibard}.  A single weak measurement results in a probability distribution similar to that depicted in Fig. \ref{fig:outcomes_weak}. For a continuous measurement signal, the result is binned into discrete time intervals $\tau$, corresponding to successive weak measurements.  Using either a stochastic master equation \cite{Dian} or Bayesian inference \cite{Vijay1, bayesianupdate1}, the state is updated after each `measurement.' 

The process by which weak measurements can be simulated is twofold. Let us take a qubit in the state $\ket{\psi} = \alpha\ket{0} + \beta\ket{1}$. After we entangle the qubit with the measurement apparatus, the result of the measurement can be given by a wave function $\ket{\psi_{out}}$ described in Eq. \eqref{eq:psiout}. To compute the respective probability densities of obtaining a particular pointer value, we take $\abs{\ket{\psi_{out}}}^2$ and observe that the outcome $\ket{0}$ is obtained with a probability $\abs{\alpha}^2$ and the outcome $\ket{1}$ is obtained with a probability $\abs{\beta}^2$. Thus, in order to simulate the result a weak quantum measurement, we must first choose a distribution based on the probabilities $\abs{\alpha}^2$ and $\abs{\beta}^2$. Using the chosen distribution, we can generate a measurement outcome. A simple model for choosing the Gaussian is a {\em biased coin toss}; this method may be more familiar as the key step used to generate binomial random variables. 
In this case, a uniformly distributed random number $r$ between 0 and 1 is generated. Given a quantum state $\rho$, the probability that $r$ is less than $\rho_{00}$ is $\abs{\alpha}^2$.  This can be implemented using the commands as shown in Fig. \ref{alg:cointoss}.

\begin{figure}[h]
\begin{algorithmic}[1]
\Require Quantum state given by $\rho = \begin{pmatrix}
\rho_{00} & \rho_{01} \\
\rho_{10} & \rho_{11}
\end{pmatrix}$ 
\Ensure $\mu \gets$ Mean of the chosen Gaussian
\Procedure{Cointoss}{$\rho_{00}$} 
\State $\mathcal{U}[a,b] \gets$ uniform random number between $[a,b]$
\State $r \gets \mathcal{U}[0,1]$ 
\If {$r \leq \rho_{00}$} $\mu \gets +1$ \Comment{For outcome $\ket{0}$}
\ElsIf {$r > \rho_{00}$} $\mu \gets -1$ \Comment{For outcome $\ket{1}$}
\EndIf
\EndProcedure
\end{algorithmic}
\caption{Algorithm for choosing a Gaussian using a biased coin toss}
\label{alg:cointoss}
\end{figure}

In actual simulations, however, the accuracy of the {\em biased coin toss} will depend on the quality of random numbers generated by $\mathcal{U}[a,b]$. Through the result of $\Call{Cointoss}{}$, we have effectively `chosen' a distribution: the $\ket{0}$ Gaussian has a mean $+1$ and the $\ket{1}$ Gaussian has a mean $-1$. 

We can use these Gaussians described by $\mathcal{N}(\mu, \sigma)$ to generate measurement outcomes $M$ following a normal distribution. Subsequently, we can use $M$ to {\em update} the state by calculating the conditional probabilities $P\!\xcmd{M}{\ket{0}}$ and $P\!\xcmd{M}{\ket{1}}$. As described earlier, these in turn give the updated state via Bayes' theorem. Given that at time $t$, $\rho_{00}(t)$ and $\rho_{11}(t)$ are simply $P(\ket{0})$ and $P(\ket{1})$ respectively, we can rewrite Eq. \eqref{eq:bayesian} using the following equations \cite{Korotkov}:
\begin{align}
\rho_{00}(t + \tau) = \frac{\rho_{00}(t) P\!\xcmd{M}{\ket{0}}}{P(M)} \label{eq:trajectory1}
\\
\intertext{and correspondingly}
\rho_{11}(t + \tau) = \frac{\rho_{11}(t) P\!\xcmd{M}{\ket{1}}}{P(M)}  
\label{eq:trajectory2}
\end{align}
On the other hand, $\rho_{01}$ is updated by
\begin{equation}
\rho_{01}(t + \tau) = \rho_{01}(t) \sqrt{\frac{\rho_{00}(t + \tau) \rho_{11}(t + \tau)}{\rho_{00}(t) \rho_{11}(t)}}
\label{eq:trajectory3}
\end{equation}

Once we have found $\rho_{01}$, we can find $\rho_{10}$ by taking the complex conjugate. As $t$ is a discrete variable here, we can set $\tau = 1$ and then treat $t$ as the index of an array for the purposes of simulation. The procedure described by Eqs. \eqref{eq:trajectory1}, \eqref{eq:trajectory2} and \eqref{eq:trajectory3} is summarized below in Fig. \ref{alg:trajectory}.

\begin{figure}[h]
\begin{algorithmic}[1]
\Require 
\Statex Initial quantum state given by $\rho(0) = \begin{pmatrix}
\rho_{00}(0) & \rho_{01}(0) \\
\rho_{10}(0) & \rho_{11}(0)
\end{pmatrix}$ 
\Statex N $\gets$ number of measurements
\Statex $\sigma \gets$ standard deviation, gives measurement strength
\Statex $\mathcal{N}(\mu, \sigma) \gets$ generates Gaussian random number
\Ensure $\rho(t)$
\Procedure{Trajectory}{}
\State $t \gets 0$
\For {$t < N$}
\State $\mu \gets \Call{Cointoss}{\rho_{00}(t)}$ \Comment{Choosing a Gaussian}
\State $M \gets \mathcal{N}(\mu, \sigma)$ \Comment{Measurement outcome} \vspace{0.2cm}
\State $P\!\xcmd{M}{\ket{0}} \gets \frac{1}{\sqrt{2\sigma^2\pi}} e^{-(M - 1)^2/2\sigma^2}$ \vspace{0.1cm}
\State $P\!\xcmd{M}{\ket{1}} \gets \frac{1}{\sqrt{2\sigma^2\pi}} e^{-(M + 1)^2/2\sigma^2}$ \vspace{0.2cm}
\State $P(M) \gets \rho_{00}(t)P\!\xcmd{M}{\ket{0}} + \rho_{11}(t)P\!\xcmd{M}{\ket{1}}$ \vspace{0.1cm}
\State $\rho_{00}(t + 1) \gets \frac{\rho_{00}(t) P\xcmd{M}{\ket{0}}}{P(M)}$ \vspace{0.1cm}
\State $\rho_{11}(t + 1) \gets \frac{\rho_{11}(t) P\xcmd{M}{\ket{1}}}{P(M)}$ \vspace{0.05cm}
\State $\rho_{01}(t + 1) \gets \rho_{01}(t) \sqrt{\frac{\rho_{00}(t + 1) \rho_{11}(t + 1)}{\rho_{00}(t) \rho_{11}(t)}}$ \vspace{0.05cm}
\State $\rho_{10}(t + 1) \gets \big(\rho_{01}(t + 1)\big)^*$
\State $t \gets t + 1$
\EndFor
\EndProcedure
\end{algorithmic}
\caption{Pseudocode for generating a quantum trajectory $\rho(t)$}
\label{alg:trajectory}
\end{figure}

\begin{figure}[h]
\subfloat[Trajectory with $\sigma = 22.36$.]{
\scalebox{1.05}{
\begin{tikzpicture}
     \begin{axis} [
     ymin=0, ymax=1,
     xmin=-10, xmax=1700,
     xlabel=t,
     xtick={250,500,750,1000,1250,1500},
     ]
\addplot[mark size=0.5,draw=blue] table [x=t, y=p11, col sep=comma] {trajectory.csv};
\addplot[mark size=0.5,draw=red] table [x=t, y=p00, col sep=comma] {trajectory.csv};
\addlegendentry{$\rho_{11}(t)$} 
\addlegendentry{$\rho_{00}(t)$} 
\end{axis}
\end{tikzpicture}}
\label{fig:trajectory}
}
\vspace{0.3cm}
\subfloat[Trajectory with $\sigma = 5$.]{
\scalebox{1.05}{
\begin{tikzpicture}
     \begin{axis} [
     ymin=0, ymax=1,
     xmin=-1, xmax=100,
     xlabel=t,
     xtick={20, 40, 60, 80,100},
     ]
\addplot[mark size=0.5,draw=blue] table [x=t, y=p11, col sep=comma] {trajectory2.csv};
\addplot[mark size=0.5,draw=red] table [x=t, y=p00, col sep=comma] {trajectory2.csv};
\addlegendentry{$\rho_{11}(t)$} 
\addlegendentry{$\rho_{00}(t)$} 
\end{axis}
\end{tikzpicture}}
\label{fig:trajectory2}}
\caption{Sample quantum trajectories of a qubit set to initial state $\rho_1$.}
\end{figure}
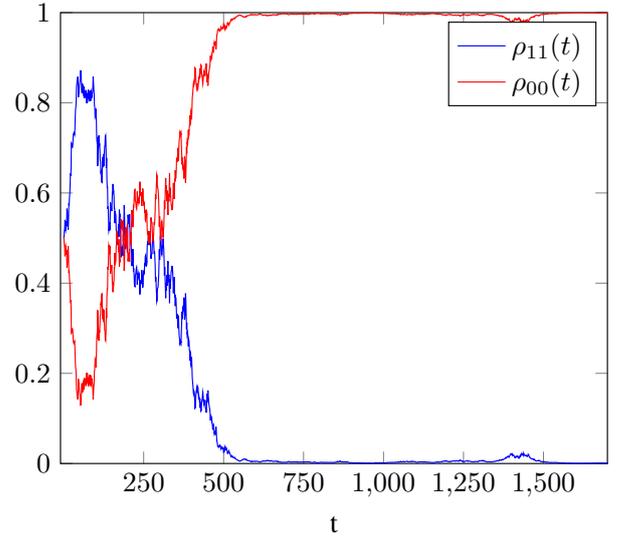
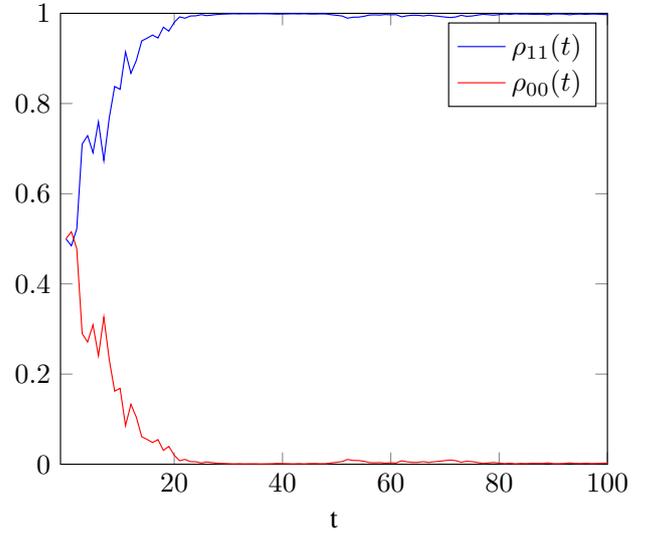

In $\Call{Trajectory}{}$, as with in $\Call{Cointoss}{}$, the quality of the random number generators will directly influence the accuracy of the method. (A convenient method for generating Gaussian random numbers is to transform a $\mathcal{U}[-1,1]$ distribution into a $\mathcal{N}(0, 1)$ distribution. The Gaussian can then be adjusted to a specified $\mu$ and $\sigma$ by taking $\sigma\times \mathcal{N}(0, 1) + \mu$. This is known as the Marsaglia method - for more details see Ref. \cite{marsaglia}). \\ As the $\rho_{00}$ and $\rho_{11}$ must always sum to 1, the trajectory $\rho_{11}(t) = 1 - \rho_{00}(t)$.
Figures \ref{fig:trajectory} and \ref{fig:trajectory2} show two such trajectories for a qubit initialized to the state $\rho_1 = \begin{pmatrix}
0.5 & 0.5 \\
0.5 & 0.5
\end{pmatrix}$. 

In both examples above $N = 100,000$. However, in the case of Fig. \ref{fig:trajectory}, we have (arbitrarily) $\sigma = 22.36$, implying a relatively weak measurement. Thus, here, the system `collapses' into one of its eigenstates after approximately 600 time steps (i.e. 600 measurements). As $\rho_{00}(t) \to 1$, it can be said with near certainty that the system eventually collapses into the $\ket{0}$ state. Conversely, in Fig. \ref{fig:trajectory2}, the system collapses into the $\ket{1}$ eigenstate, as indicated by the value of $\rho_{11}(t) \to 1$. In this case, however, the measurement is relatively stronger, with $\sigma = 5$. Consequently, the `collapse' occurs after only 25 time steps: that is to say, after 25 measurements the system is almost certainly in the $\ket{1}$ state. Of course, in order to arrive at a general conclusion about the average `collapse' time, one must average over many trajectories. As the measurement strength increases, this average collapse time will reduce. Intuitively this makes sense, as for small values of  $\sigma$, the state would immediately collapse into one eigenstate upon measurement. Likewise, by observing Figs. \ref{fig:trajectory} and \ref{fig:trajectory2}, we notice that successive weak measurements eventually give the same result as a strong measurement: the wave function collapsing into one of the eigenstates.  This fact is well-supported theoretically: Oreshkov et. al. showed that any generalized measurement can be written as a series of weak measurements \cite{Oreshkov}.

\section{Method: Single-qubit Quantum State Tomography}
	
    Quantum state tomography is the process by which an unknown state is `reconstructed' from a series of measurements on the state \cite{Wu}. Of course, by performing state tomography on a known state, and comparing this to the estimated state, we can determine the efficacy of the tomography process.  As a classic example given in Refs. \cite{QCQI,D'Ariano}, imagine that Alice prepares an ensemble of qubits and gives it to Bob to observe. Bob, skeptical of Alice's description, can perform quantum state tomography on the ensemble to determine the true state of the qubits. In an ideal scenario, Bob would be able to carry out measurements on an infinite number of copies of the qubit. However, in practice, experiments are limited by a fixed ensemble size.

      From literature, we know that projective measurements outperform weak measurements for state reconstruction in actual experiments, where ensemble sizes are large (with over thousands of copies of the qubit state). However, Das and Arvind raise the question of whether weak measurements  possess an advantage for small ensemble sizes. This is prompted by the possibility of recycling states for subsequent measurements. Das and Arvind propose the following scheme: weak measurements are performed on the $z$ and $x$ Pauli components of spin, followed by a projective measurement along the $y$ component. This scheme is tested, and it is shown that under certain conditions, weak measurements can outperform projective measurements. Das and Arvind claim this scheme offers a ``novel way of extracting information from a quantum system'', as the two weak measurements extract information without greatly disturbing the state.
    
\begin{figure}[h]
    \centering
    \includegraphics[width=0.75\columnwidth]{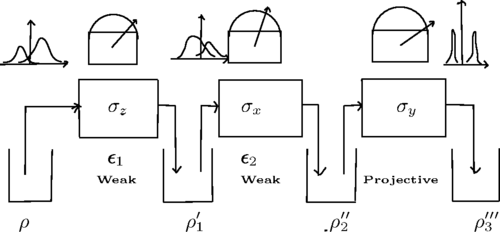}
    \caption{Depiction of the Das-Arvind protocol. Reprinted from Ref. \cite{Arvind}}
    \label{fig:arvindprotocol}
    \end{figure}

    The protocol considers a fixed ensemble of identical qubits in the state $\rho$. On this ensemble, a $\sigma_z$ measurement is carried out with a coupling strength $\epsilon_1$. For each qubit the meter reading is observed. Due to the associated backaction of the measurements, the state of the qubits is changed, denoted by the density matrix $\rho_1$. Next, a $\sigma_x$ measurement is performed with coupling strength $\epsilon_2$, following the same process as for $\sigma_z$. This gives an updated state $\rho_2$. Lastly, a projective $\sigma_y$ measurement is carried out on each of the qubits in the ensemble. This subroutine is repeated a fixed number of times, and the results are averaged in order to avoid statistical error. A summary of this subroutine is given in Fig. \ref{fig:arvindprotocol}. In the first two measurements, the Gaussians  overlap, indicating that the measurement strength is weak. These are similar to the illustration in Fig. \ref{fig:outcomes_weak}. In the $\sigma_y$ measurements, however, the Gaussians are well separated. By computing the expectation values for $\sigma_x$, $\sigma_y$, and $\sigma_z$, we arrive at the Bloch sphere vector $(x,y,z)$ of the state.    
    
When the Gaussians overlap significantly, and the system has similar probability of being in either eigenstate, it becomes difficult to ascribe a meter reading to its associated eigenstate. Das and Arvind address this by introducing a {\em discard parameter}. This is defined as a region in between the two Gaussians of width $2a$. Any meter reading that falls in this region is not considered; however, the state is preserved and recycled for subsequent measurements. Meter readings that do lie outside of this `discard region' are binned into two categories. For example, in the case of the $\sigma_x$ measurement, readings to the right of this region indicate $\sigma_x$ has a value $+1$, whereas readings to the left indicate a value $-1$.    
    
    This study seeks to reproduce their results. As we will see in Section IV, it is surprising that a weak measurement scheme should outperform a protocol based on projective measurements. Our first step involves recreating the procedure used by Das and Arvind. In their scheme, Das and Arvind calculate the probability density function using the reduced density matrix of the measurement apparatus, which itself is obtained by taking the partial trace of the state. We, however, make use of Bayesian inference to update the density matrices. In theory, this is functionally equivalent and does \textit{not} degrade the quality of the state tomography. Furthermore, we work with the assumption that the strength of both weak measurements is equal -- that is, $\epsilon_1 = \epsilon_2 = \epsilon$. Imagine now that the ensemble of $n$ qubits is prepared in a state given by 
    \begin{equation}
    \rho = \rho_{00}\ket{0}\bra{0} + \rho_{01}\ket{0}\bra{1} + \rho_{10}\ket{1}\bra{1} + \rho_{11}\ket{1}\bra{1} 
    \end{equation}

Referring to Eq. \ref{eq:densitymatrix2}, we can compute the coordinates of the Bloch sphere vector $(x,y,z)$ as 
\begin{align}
x &= \rho_{01} + \rho_{10}\\
y &= i(\rho_{01} - \rho_{10}) \\
z &= \rho_{00} - \rho_{11} 
\end{align}

On each member of the ensemble, three measurements are performed. For $\sigma_z$ the measurement outcome $M_z$ is recorded and used to update the state using Bayesian inference (the process is the same used as in \Call{Cointoss}{} and \Call{Trajectory}{}). However, the value of $M_z$ may fall inside the discard region for some of the $n$ qubits in the ensemble. Thus, we define the number of `valid measurement outcomes' $C_{\sigma z}$. If a measurement is `valid', then we increment a sum $S_{\sigma z}$ by $+1$ for values of $M_z$ to the right of the discard region, and by $-1$ for values of $M_z$ to the left of the discard region. The expectation value $\langle \sigma_z\rangle$ is then given by
\begin{equation}
\langle\sigma_z\rangle = \frac{S_{\sigma z}}{C_{\sigma z}}
\end{equation}

Next, before making a measurement on $\sigma_x$, the qubit must be rotated to the correct orientation; this is because in physical devices, and in this implementation, measurements can only be made along one axis. The matrices for rotation by an angle $\theta$ about the x- and y-axes are given in Ref. \cite{QCQI} as,

\begin{equation*}
R_x(\theta) = 
\begin{pmatrix}
\cos\frac{\theta}{2} & -i\sin\frac{\theta}{2} \\
-i\sin\frac{\theta}{2} & \cos\frac{\theta}{2}
\end{pmatrix} \quad
R_y(\theta) = 
\begin{pmatrix}
\cos\frac{\theta}{2} & -\sin\frac{\theta}{2} \\
\sin\frac{\theta}{2} & \cos\frac{\theta}{2}
\end{pmatrix}
\end{equation*} 

All this while, we have considered our measurement device to be aligned with $\sigma_z$. Thus, we must rotate the updated state $\rho_1$  around the y-axis by $\theta =-90^\circ$ such that the qubit x-axis shifts to the position of the z-axis. This is carried out by
\begin{equation}
\rho_1^{x-\text{basis}} = R_y\big(-90^\circ\big) \rho_1 R_y\big(-90^\circ\big)^\dagger
\end{equation}
where $^\dagger$ is the conjugate transpose matrix operation. Using this state $\rho_1^{x-\text{basis}}$ we perform a measurement on $\sigma_x$ through the same process used to measure $\sigma_z$. After generating a measurement outcome $M_x$, the state is updated, giving a new density matrix $\rho_2^{x-\text{basis}}$. Likewise, the number of valid measurements is noted and the expectation value is given by 
\begin{equation}
\langle\sigma_x\rangle = \big(\frac{S_{\sigma x}}{C_{\sigma x}}\big) e^{\epsilon/2}
\end{equation}
Here, a correction factor (as given in Ref. \cite{Arvind}) is introduced to compensate for the backaction on the system. This factor is dependent on measurement strength parameter $\epsilon$. Finally, in order to make a $\sigma_y$ measurement, the qubit must be rotated into the y-basis. First, the qubit state is rotated back into the original (z-) basis by $\rho_2 = R_y\big(-90^\circ\big)^\dagger \rho_2^{x-\text{basis}}  R_y\big(-90^\circ\big)$. Next, the qubit state is rotated by $\theta = 90^\circ$ about the x-axis, so that
\begin{equation}
\rho_2^{y-\text{basis}} = R_x\big(90^\circ\big) \rho_2 R_x\big(90^\circ\big)^\dagger
\end{equation}
By performing a \Call{Cointoss}{} operation with this state $\rho_2^{y-\text{basis}}$ we obtain either $+1$ or $-1$. These values are added to the sum $S_{\sigma y}$. As there is no discard parameter for a projective measurement, the number of `valid' measurements will simply be the number of qubits in the ensemble, $n$. Here, the expectation value is given by
\begin{equation}
\langle\sigma_y\rangle = \bigg[2\Big(\frac{S_{\sigma y}}{n}\Big) - 1 \bigg] e^\epsilon
\label{eq:sigmay}
\end{equation}
Once again, a correction factor accounts for the backaction of the two prior measurements. A full proof of the calculation of $\langle\sigma_y\rangle$ is given in Appendix A, but the intuition behind this result comes from realizing that the ratio $S_{\sigma y}/n$ is simply $P(\sigma_y;+)$. These expectation values directly give us the coordinates of the Bloch sphere vector -- that is, $\langle\sigma_x\rangle = x_{est}$, $\langle\sigma_y\rangle = y_{est}$, and $\langle\sigma_z\rangle = z_{est}$.

For the scheme of projective measurements, on the other hand, the ensemble is divided into three parts. Each part is first rotated into the correct basis, before a \Call{Cointoss}{} operation is performed. The results of these measurements are used to calculate the expectation values, using the calculation made in Eq. \eqref{eq:sigmay} for all three components, except without the correction factor. From here, we arrive at $x_{est}$, $y_{est}$, and $z_{est}$, and can compare the results for both measurement schemes. In order to quantify the comparison, we introduce the fidelity $f$ as the distance between the actual and estimated Bloch sphere vectors. This measure is given by Das and Arvind as:
\begin{equation}
f = 1 - \bigg[(x - x_{est})^2 + (y - y_{est})^2 + (y - y_{est})^2\bigg]
\end{equation}

We repeat this process to obtain average trends of the tomographic process. In the interest of brevity, the algorithm summarizing the process above is described in Appendix B. Nonetheless, we list here the parameters for the simulation in Table \ref{tab:parameters}. 
\begin{table}[hp]%
\caption{List of parameters used in the simulation}
\label{tab:parameters}\centering %
\vspace{-0.1cm}
\begin{tabular}{>{\centering\arraybackslash} m{1.2cm} >{\centering\arraybackslash} m{1.2cm} >{\centering\arraybackslash} m{4.65cm} }
\toprule %
Simulation & Physical & Description \\
Parameter & Significance & of parameter \\\midrule %
{\tt rho} & $\rho$ & Quantum state being estimated \\
{\tt ensemble} & $n$ & Number of qubits in the ensemble \\
{\tt epsilon} & $\epsilon$ & Gives the easurement strength \\
{\tt sigma} & $\sigma$ & Standard deviation of Gaussians ($1/\sqrt{\epsilon}$) \\
 {\tt N} & - & Number of reiterations of protocol \\
{\tt a} & $a$ & Discard parameter \\\bottomrule
\end {tabular}
\end{table}

\section{Results}

\subsection{Example of a randomly-generated state}
We first test the scheme on a randomly generated state  $\rho_A$, given by
\begin{equation}
\rho_A = \frac{1}{2}\begin{pmatrix}
1.399 & -0.385 + 0.042i \\
-0.385 - 0.042i & 0.601
\end{pmatrix}
\end{equation}

This quantum state is used as a test case by Das and Arvind (see Ref. \cite{Arvind}). Thus, testing the protocol allows us to directly compare our results. As with Das and Arvind, we take an ensemble size of 30. However, we find that repeating the process only 10000 times (as done in Ref. \cite{Arvind}) gives rise to large statistical fluctuations. Thus, here, we instead use  100000 iterations to avoid such difficulties. In each run, the fidelity $f$ is calculated. We then calculate the mean fidelity $\bar{f}$ and standard deviation in fidelity $\sigma_{\bar{f}}$ for that particular measurement strength $\epsilon$. By varying $\epsilon$, we are able to reproduce the results noted by Das and Arvind for the state $\rho_A$. 

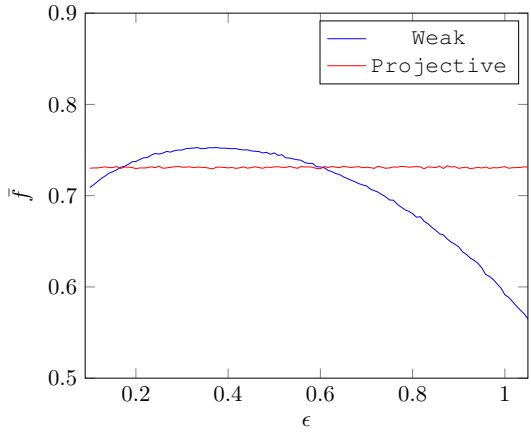
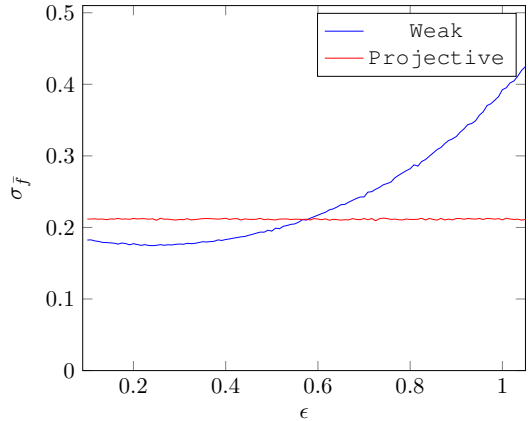
\begin{figure}[h]
\centering
\subfloat[Plot of mean fidelity $\bar{f}$ vs. $\epsilon$]{
\scalebox{0.85}{
\begin{tikzpicture}
     \begin{axis} [
     ymin=0.5, ymax=0.9,
     xmin=0.09, xmax=1.05,
     xlabel=$\epsilon$,
     ylabel=$\bar{f}$,
     ]
\addplot[mark size=0.5,draw=blue] table [x=epsilon, y=fidelity, col sep=comma] {rhoA_weak.csv};
\addplot[mark size=0.5,draw=red] table [x=epsilon, y=fidelity, col sep=comma] {rhoA_projective.csv};
\addlegendentry{\tt Weak}
\addlegendentry{\tt Projective}
\end{axis}
\end{tikzpicture}}}
\vspace{0.3cm}
\subfloat[Plot of standard deviation in fidelity $\sigma_{\bar{f}}$ vs. $\epsilon$]{
\scalebox{0.85}{
\begin{tikzpicture}
     \begin{axis} [
     ymin=0, ymax=0.51,
     xmin=0.09, xmax=1.05,
     xlabel=$\epsilon$,
     ylabel=$\sigma_{\bar{f}}$,
     ]
\addplot[mark size=0.5,draw=blue] table [x=epsilon, y=std_dev, col sep=comma] {rhoA_weak.csv};
\addplot[mark size=0.5,draw=red] table [x=epsilon, y=std_dev, col sep=comma] {rhoA_projective.csv};
\addlegendentry{\tt Weak}
\addlegendentry{\tt Projective}
\end{axis}
\end{tikzpicture}}}
\\
\caption{Plots of mean fidelity (a) and standard deviation in fidelity (b) versus $\epsilon$ for weak (blue) and projective (red) measurement schemes. The ensemble of 30 qubits is prepared in the state $\rho_A$, and the discard parameter $a = 0$.}
\label{fig:rhoAplot}
\end{figure}

Through an analysis of Fig. \ref{fig:rhoAplot}, we see that the mean fidelity for the weak measurement protocol peaks at $\epsilon \approx 0.4083$. In this case, the weak measurement scheme offers an improvement over the projective scheme. We clarify that the projective measurement scheme, based on a series of biased coin tosses, results in a roughly constant fidelity. The reason for the weak measurement scheme outperforming the projective scheme, in this case, is not obvious; the former consists of a series of projective $\sigma_y$ measurements, while the latter makes projective measurements on $\sigma_x$, $\sigma_y$, and $\sigma_z$. Likewise, it is also unclear why the weak measurement scheme displays the trend that is does. The value of $\epsilon$ for which the data peaks, in the case of $\rho_A$, corresponds to a measurement strength that is neither extremely weak nor extremely strong. One hypothesis offered by Das and Arvind is that for very weak measurements, the Gaussians are too wide to be able to determine which eigenstate the pointer value belongs to, whereas for very strong measurements, the state collapses after only one measurement. This thinking is in accordance with Lundeen et. al., who claim that weak measurements must be carried out ``in a gentle way'' \cite{Lundeen}. Nonetheless, having reproduced the results of Das and Arvind, we are encouraged to continue the analysis.

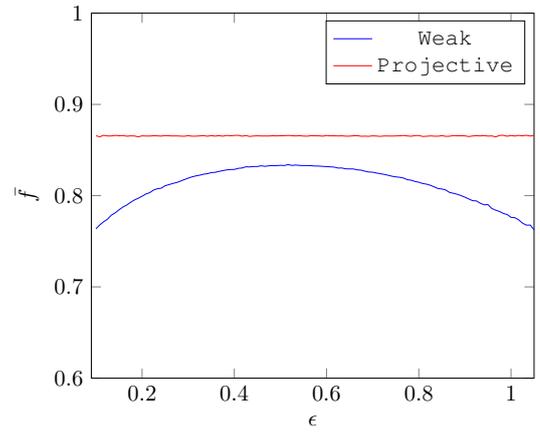
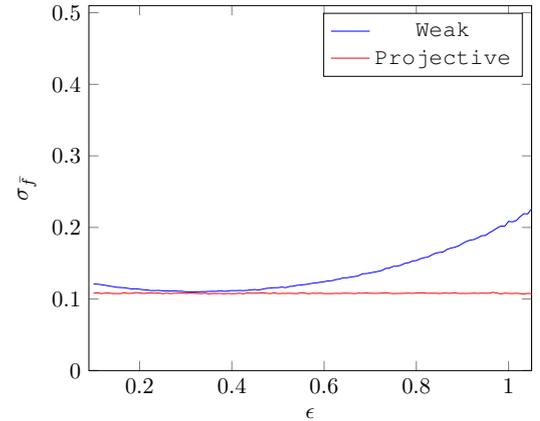
\begin{figure}[h]
\centering
\subfloat[Plot of mean fidelity $\bar{f}$ vs. $\epsilon$]{
\scalebox{0.85}{
\begin{tikzpicture}
     \begin{axis} [
     ymin=0.6, ymax=1,
     xmin=0.09, xmax=1.05,
     xlabel=$\epsilon$,
     ylabel=$\bar{f}$,
     ]
\addplot[mark size=0.5,draw=blue] table [x=epsilon, y=fidelity, col sep=comma] {rhoA_weak_60.csv};
\addplot[mark size=0.5,draw=red] table [x=epsilon, y=fidelity, col sep=comma] {rhoA_projective_60.csv};
\addlegendentry{\tt Weak}
\addlegendentry{\tt Projective}
\end{axis}
\end{tikzpicture}}}
\vspace{0.3cm}
\subfloat[Plot of standard deviation in fidelity $\sigma_{\bar{f}}$ vs. $\epsilon$]{
\scalebox{0.85}{
\begin{tikzpicture}
     \begin{axis} [
     ymin=0, ymax=0.51,
     xmin=0.09, xmax=1.05,
     xlabel=$\epsilon$,
     ylabel=$\sigma_{\bar{f}}$,
     ]
\addplot[mark size=0.5,draw=blue] table [x=epsilon, y=std_dev, col sep=comma] {rhoA_weak_60.csv};
\addplot[mark size=0.5,draw=red] table [x=epsilon, y=std_dev, col sep=comma] {rhoA_projective_60.csv};
\addlegendentry{\tt Weak}
\addlegendentry{\tt Projective}
\end{axis}
\end{tikzpicture}}}
\\
\caption{Plots of mean fidelity (a) and standard deviation in fidelity (b) versus $\epsilon$ for weak (blue) and projective (red) measurement schemes. The ensemble of 60 qubits is prepared in the state $\rho_A$, and the discard parameter $a = 0$.}
\label{fig:rhoAplot60}
\end{figure}

We next repeat the protocol on the state $\rho_A$, with an ensemble size of 60 qubits. As seen in Fig. \ref{fig:rhoAplot60}, this increases the fidelity of the projective scheme. The reason for this is not difficult to see: more coin tosses will give a better estimate of the associated probabilities, and thus a more accurate reconstruction of the state. However, we see that on average the weak measurement scheme does {\em not} outperform the projective scheme, for a larger ensemble. We can, however, compare the results of Fig. \ref{fig:rhoAplot60} to those in \ref{fig:rhoAplot} to see that increasing the ensemble size does improve the fidelity for both schemes. However, for the projective measurement scheme this improvement is greater; Fig. \ref{fig:rhoAplot60} shows that the weak scheme peaks at $\bar{f} \approx 0.82$, whereas for the projective scheme $\bar{f} \approx 0.86$ consistently. Das and Arvind also acknowledge that their weak measurement scheme is only effective for small ensemble sizes, and thus this result is not surprising. However, it is interesting to note that by changing the ensemble size, the peak fidelity appears to occur at a different value of $\epsilon$.

\subsection{Measurement Fidelity Depends on the State}

We now extend the protocol to test different quantum states. We consider the simple case $\rho_B$, where
\begin{equation}
\rho_B = \begin{pmatrix}
0.5 & 0.5 \\
0.5 & 0.5
\end{pmatrix}
\end{equation}

The parameters of this simulation are identical to those used to measure $\rho_A$ in Fig. \ref{fig:rhoAplot}. That is, the protocol is run for 30 qubits and repeated 100000 times. We set the discard parameter to 0. 

\begin{figure}[h]
\centering
\subfloat[Plot of mean fidelity $\bar{f}$ vs. $\epsilon$]{
\scalebox{0.85}{
\begin{tikzpicture}
     \begin{axis} [
     ymin=0.3, ymax=1,
     xmin=0.09, xmax=1.05,
     xlabel=$\epsilon$,
     ylabel=$\bar{f}$,
     ]
\addplot[mark size=0.5,draw=blue] table [x=epsilon, y=fidelity, col sep=comma] {rhoB_weak.csv};
\addplot[mark size=0.5,draw=red] table [x=epsilon, y=fidelity, col sep=comma] {rhoB_projective.csv};
\addlegendentry{\tt Weak}
\addlegendentry{\tt Projective}
\end{axis}
\end{tikzpicture}}}
\vspace{0.3cm}
\subfloat[Plot of standard deviation in fidelity $\sigma_{\bar{f}}$ vs. $\epsilon$]{
\scalebox{0.85}{
\begin{tikzpicture}
     \begin{axis} [
     ymin=0, ymax=0.51,
     xmin=0.09, xmax=1.05,
     xlabel=$\epsilon$,
     ylabel=$\sigma_{\bar{f}}$,
     ]
\addplot[mark size=0.5,draw=blue] table [x=epsilon, y=std_dev, col sep=comma] {rhoB_weak.csv};
\addplot[mark size=0.5,draw=red] table [x=epsilon, y=std_dev, col sep=comma] {rhoB_projective.csv};
\addlegendentry{\tt Weak}
\addlegendentry{\tt Projective}
\end{axis}
\end{tikzpicture}}}
\\
\caption{Plots of mean fidelity (a) and standard deviation in fidelity (b) versus $\epsilon$ for weak (blue) and projective (red) measurement schemes. The ensemble of 30 qubits is prepared in the state $\rho_B$, and the discard parameter $a = 0$.}
\label{fig:rhoBplot}
\end{figure}
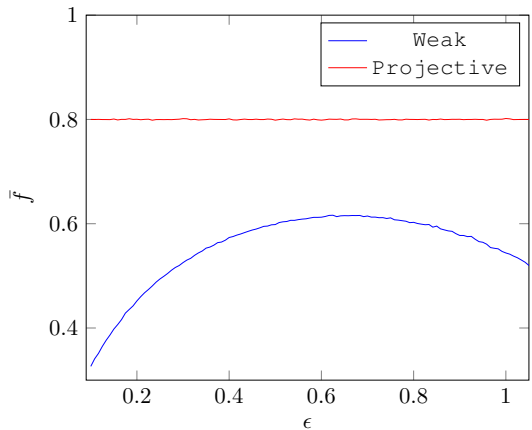
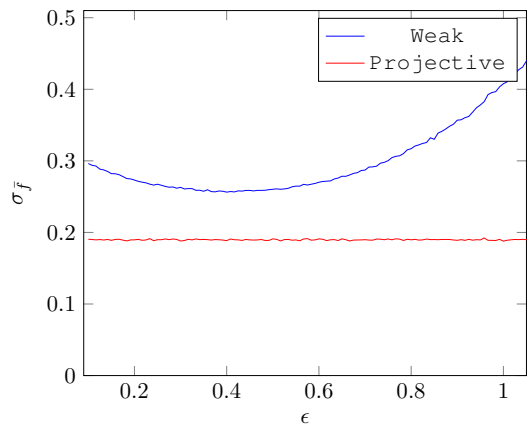

On Fig. \ref{fig:rhoBplot}, we see that the weak measurement scheme performs quite poorly compared to the projective scheme. Let us emphasize that the simulation parameters here are identical to those used to measure state $\rho_A$ in Fig. \ref{fig:rhoAplot}. Here, the projective measurement scheme has a mean fidelity $\bar{f} \approx 0.80$, whereas the weak measurement scheme peaks at $\bar{f} \approx 0.62$. Likewise, it is interesting that this peak should occur at $\epsilon = 0.625$, as compared to $\epsilon = 0.408$ for the state $\rho_A$.

\section{Discussion and Summary}

In this study, we have been able to reproduce the results of Das and Arvind. In particular, we see that for certain states, the Das-Arvind weak measurement protocol offers slight advantages in mean measurement fidelity. We further verify that the phenomenon observed is dependent on ensemble-size. By running the simulation for larger ensembles, we see that both the Das-Arvind scheme and projective measurement scheme perform better with more qubits -- however, the improvement is more pronounced for the projective measurement case. Clearly, as Das and Arvind themselves take note of, the Das-Arvind protocol only performs better for small ensembles. Finally, we see that the protocol does not offer improvements for all states, and that the fidelity of the process is state dependent.

At this stage it is worth calling into question the `binning' process, by which meter readings to the left of the discard region are interpreted to be -1 and those to the right are interpreted as +1. We note, interestingly, that if this interpretation is not made (and therefore if the pointer value is added when calculating the sum instead of simply +1 or -1) then the Das-Arvind protocol performs better. Indeed as per Ref. \cite{Gross}, this binning process ``degrades tomographic performance.''

The results of Das and Arvind, reproduced here, raise a number of questions for which states the Das-Arvind protocol offers improvements, and the optimal ensemble sizes in each case. Das and Arvind believe that the two weak measurements followed by a projective measurement allow for more information to be extracted from a qubit than from a projective measurement alone. However, we know from Busch's theorem that any information gained from a quantum system necessarily disturbs the state \cite{Busch}. The limit on the amount of information that can be extracted is given in Ref. \cite{Massar}. 

Instead, we try to explain the trend observed here by placing the Das-Arvind protocol in the context of a more recent work by Gross et. al. \cite{Gross} In their analysis of the Das-Arvind protocol, the authors clarify that the projective measurement scheme used here and in \cite{Arvind} is a type of a mutually unbiased basis (MUB) scheme, where three orthogonal measurement bases (i.e x- y- and z-axes) are chosen. Gross et. al. show, however, that the MUB scheme is not the most efficient scheme for projective measurements. Indeed, the most efficient scheme involves choosing a random measurement basis from a uniformly distributed set of axes. This is known as a Haar-invariant measurement, and it falls under the general classification of one-dimensional orthogonal projective measurements (ODOPs). 

Intuitively, it is not difficult to see why ODOPs provide higher fidelity measurements. Imagine a pure state pointing in the $\ket{\sigma_z;+}$ direction; then the $\sigma_z$ measurements will be highly accurate, whereas the $\sigma_x$ and $\sigma_y$ measurements give the correct result only 50\% of the time (i.e relatively inaccurate). Likewise, generally, for a fixed-basis measurement scheme, the quality of state reconstruction will depend on the choice of measurement basis. This makes sense given that we observed the Das-Arvind protocol to be state dependent -- the choice of basis is optimal for some states, but not for others. ODOPs on the other hand would overcome this by randomly sampling bases.

Gross et. al. show that the Das-Arvind protocol does not outperform their ODOP-based protocol. Through their analysis it is shown that the first two weak measurements cause random backaction on the state, and result in a distribution around the Bloch sphere from which the final projective measurement is sampled. For the optimum value of $\epsilon$, this results in a uniform distribution. Under these conditions the final projective measurement in the Das-Arvind scheme is essentially random, approximating an ODOP. Thus, although more information is extracted using the Das-Arvind protocol than simply a $\sigma_y$ projective measurement alone, this is not due to any improvements offered by weak measurements. Instead, the weak measurements in this scheme only serve to randomly rotate the state, an effect that could be achieved by using a randomly-selected unitary (as is used in ODOPs). Gross et. al. thus conclude that the Das-Arvind protocol does not offer any improvements over an ODOP based scheme. At the present stage in our understanding, our data supports this claim, as the Das-Arvind protocol offers only marginal improvements under certain specific conditions.

\section{Future Work}

In the light of \cite{Gross}, we hope to extend this analysis of weak and projective measurement schemes to include other types of generalized measurements. In doing so, we hope to verify the claims made by Gross et. al., and test the efficiency of a Haar-invariant ODOP-based scheme. We will also work towards comparing the efficacy of these various tomographic processes experimentally. 

Despite some dispute about the claims of several weak measurement tomography based schemes, the notion of the weak measurement maintains its theoretical and practical significance in quantum computing. From Refs. \cite{Busch} and \cite{Massar} we know that there is no way around the uncertainty-disturbance relations, for both weak and strong measurements. Nonetheless, weak measurements may find use in so-called adaptive measurements, where the measurement bases are continuously updated based on the results of a series of weak measurements. Thus, testing such protocols and comparing to the non-adaptive schemes presented here is a further goal for us.

\section{Acknowledgements}

The author would like to thank Dr. Rajamani Vijayaraghavan of the Quantum Measurement and Control Laboratory at the Tata Institute of Fundamental Research (TIFR) for his mentorship and support, as well as Malay Singh and Soumen Das for their invaluable guidance and advice.

    % Now we need a bibliography:

\appendices
\section{How to calculate $\langle\sigma_y\rangle$}
Here, we continue from Section III, offering a short proof for the result in Eq. \eqref{eq:sigmay}. We first imagine trying to find $\langle\sigma_z\rangle$ (yes, in the z-direction) for an arbitrary qubit state $\rho$. This can be calculated by the trace of $\rho$ times the $\sigma_z$ Pauli matrix, or
\begin{equation}
\langle\sigma_z\rangle = Tr(\rho \sigma_z) = \rho_{00} - \rho_{11}
= \rho_{00} - (1 - \rho_{00})
\end{equation}
As here $\rho_{00} = P(\ket{0})$, we can rewrite this as 
\begin{equation}
\langle\sigma_z\rangle = 2P(\ket{0}) - 1
\label{eq:sigmay_proof}
\end{equation}
In Section III, with Eq. \eqref{eq:sigmay}, we rotated the qubit y-axis to the position of the z-axis. Thus, the process used here can be effectively used to calculate $\langle\sigma_y\rangle$ (as we are measuring along the z-axis of the rotated state, corresponding to the y-axis of the original state). Replacing $\langle\sigma_z\rangle$ with $\langle\sigma_y\rangle$ and $P(\ket{0}) = P(\sigma_z;+)$ with $P(\sigma_y;+)$ in Eq. \eqref{eq:sigmay_proof} yields Eq. \eqref{eq:sigmay}. The `correction factor' can then be added by hand.

\section{An algorithm to implement the Das-Arvind protocol}
We present parts of the algorithm (written using C) that was used to generate our results. In order to run this code using the gcc compiler, the following command-line argument must be provided: $\texttt{gcc -o FILENAME FILENAME.c -lm -lbsd}$. These linker commands are essential, as we make use of the following libraries: {\tt <math.h>, <complex.h>, and <stdlib.h>}

In order to work with the operators of quantum mechanics, we define a new data type to handle $2\times2$ matrices. This is done implemented using a {\tt struct}: 

\begin{lstlisting}[
    basicstyle=\small,]
typedef struct {
    double complex p00, p01, p10, p11;
} matrix;
\end{lstlisting}

All rotation matrices, and quantum states are defined using this data type. For the latter, note that the numbers {\tt p00, p01, p10} and {\tt p11} correspond to the matrix elements $\rho_{00}$, $\rho_{01}$, $\rho_{10}$, and $\rho_{11}$. Using our new data type, we also define a simple {\tt multiply()} function to multiply the corresponding matrix elements. This is necessary to perform the necessary qubit rotations. 

For any given run, we perform $\sigma_x$, $\sigma_y$, and $\sigma_z$ measurements; in our implementation, we simply loop the process up till the number of qubits, {\tt ensemble}. For an input state {\tt rho}, we first compute the Bloch sphere coordinates using: 

\begin{lstlisting}[
    basicstyle=\small,]
z_actual = rho.p00 - rho.p11;
y_actual = (1I)*(rho.p01 - rho.p10);
x_actual = rho.p01 + rho.p10;
\end{lstlisting}

where the constant {\tt I} is imported from {\tt <complex.h>} and represents $i = \sqrt{-1}$. 
Next we simulate a weak measurement. As mentioned in Section II, this is done using a combination of a biased coin toss and a Gaussian random number to generate a measurement outcome {\tt M}. 

\begin{lstlisting}[
    basicstyle=\small,]
double coin = cointoss(creal(rho.p00));
if (coin == 0)
   M = gaussrand(sigma, mean0);
else if (coin == 1)
   M = gaussrand(sigma, mean1);
\end{lstlisting}

The {\tt cointoss()} function is defined exactly like \Call{Cointoss}{} from Section II. However, we must convert from a {\tt complex double} data type to a {\tt double}. To generate uniform random numbers, we use the {\tt arc4random()} function from {\tt <stdlib.h>}. It produces random numbers of sufficient quality for our simulation: 

\begin{lstlisting}[
    basicstyle=\small,]
double cointoss(double p) {
    double x = arc4random() / (RAND_MAX + 1.0); 
    //uniform random number between -1 and 1
    x = .5*(x + 1); 
    //adjusts the distribution to be over [0,1]
    if (x <= p)
        return 0;
    else if (x > p)
        return 1;
}
\end{lstlisting}

Using the result of the {\tt cointoss()} function, we randomly select a measurement outcome from a Gaussian distribution. This is done with the {\tt gaussrand()} function, which uses the Marsaglia method \cite{marsaglia} to generate normally distributed variables with mean 0 and standard deviation 1. In our definition {\tt gaussrand()} takes two arguments: {\tt sigma} and the mean of the distribution ({\tt mean0 = +1} and {\tt mean1 = -1}). It adjusts the Gaussian distribution by multiplying by {\tt sigma} and adding the mean. 
 
\begin{lstlisting}[
    basicstyle=\small,]
double gaussrand(double sigma, double mean) { 
  static double V1, V2, S;
  static int phase = 0;
  double X;
  double U1, U2;
    
  if (phase == 0) {
   do {
     U1 = arc4random() / (RAND_MAX + 1.0);
	 U2 = arc4random() / (RAND_MAX + 1.0);
	 V1 = 2 * U1 - 1;
	 V2 = 2 * U2 - 1;
	 S = V1 * V1 + V2 * V2;
	 } while(S >= 1 || S == 0);

	 X = V1 * sqrt(-2 * log(S) / S);
  } else
	 X = V2 * sqrt(-2 * log(S) / S);
  phase = 1 - phase;
  return (double) ((X * sigma) + mean);
\end{lstlisting}

Once we have generated a measurement outcome {\tt M}, we can update the state using Bayesian inference. The procedure for this is given in the \Call{Trajectory}{} function in Section II. However, we add the following condition which bins the results according to whether {\tt M} lies inside the discard region of width $2a$: 
\begin{lstlisting}[
    basicstyle=\small,]
if (M >= a) {
   sum_x += 1.0;
   count_x += 1.0;
} else if (M <= -a) {
   sum_x += -1.0;
   count_x += 1.0;}
\end{lstlisting}

As listed in Section III, the expectation values can be calculated using these counts. We implement this as follows:

\begin{lstlisting}[
    basicstyle=\small,]
ex_z[j] = (sum_z / count_z);
ex_x[j] = (sum_x / count_x)*pow(E, (epsilon/2));
ex_y[j] = (2*(count_y / ensemble) - 1)
				*pow(E, epsilon);
double delz = z_actual - ex_z[j];
double delx = x_actual - ex_x[j];
double dely = y_actual - ex_y[j];
fidelity[j] = 1 - (pow(delx, 2) + pow(dely, 2) 
				+ pow(delz, 2));
\end{lstlisting}

Here we are calculating the fidelity for the {\tt j}th iteration. We run an outer {\tt for} loop up till {\tt N} (the number of reiterations), we can calculate the mean fidelity by taking the average value of the array {\tt fidelity[]}. Another {\tt for} loop can then be used to repeat this process for different values of {\tt epsilon} (and hence of {\tt sigma}). If we then plot mean fidelity against {\tt epsilon}, we obtain the figures seen in Section IV. 

% Your document ends here!
\end{document}